\documentclass[12pt]{article}

\usepackage{amssymb}
\usepackage{graphicx}

\begin{document}

\title{Entanglement and Localization of a Two-Mode Bose-Einstein Condensate}

\author{Citlali P\'erez-Campos, Jos\'e Ra\'ul Gonz\'alez-Alonso, \\
Octavio Casta\~nos, and Ram\'on L\'opez-Pe\~na\\
\small{Instituto de Ciencias Nucleares, }\\
\small{Universidad Nacional Aut\'onoma de M\'exico,}\\   \small{Apartado Postal 70-543, 
04510 M\'exico, DF, M\'exico.}}

%\address{Instituto de Ciencias Nucleares, 
%Universidad Nacional Aut\'onoma de M\'exico,\\   Apartado Postal 70-543, 
%04510 M\'exico, DF, M\'exico.}

\date{}

 \maketitle
 
\begin{abstract}
A simple second quantization model is used to describe a two-mode Bose-Einstein
condensate (BEC), which can be written in terms of the generators of a $SU(2)$ algebra with three parameters. We study the behaviour of the entanglement entropy and localization of the system in the parameter space of the model. The phase transitions in the parameter space are determined by means of the coherent state formalism and the catastrophe theory, which besides let us get the best variational state that reproduces the ground state energy. This semiclassical method let us organize the energy spectrum in regions where there are crossings and anticrossings.
The ground state of the two-mode BEC, depending on the values of the interaction strengths, is dominated by a single Dicke state, a spin collective coherent state, or a superposition of two spin collective coherent states. The entanglement entropy is determined for two recently proposed partitions of the two-mode BEC that are called separation by boxes and separation by modes of the atoms. The entanglement entropy in the boxes partition is strongly correlated to the properties of localization in phase space of the model, which is given by the evaluation of the second moment of the Husimi function.  To compare the fitness of the trial wavefunction its overlap with the exact quantum solution is evaluated.  The entanglement entropy for both partitions, the overlap and localization properties of the system get singular values along the separatrix of the two-mode BEC, which indicates the phase transitions which remain in the thermodynamical limit, in the parameter space.

\end{abstract}

PACS number(s): 03.75.Gg; 03.65Ud; 05.30Jp; 03.67Mn

\section{Introduction}

The BEC phenomena has a long history and it starts when A. Einstein, on the basis of a work of S. Bose devoted to the statistical description of the quanta of light, predicted that when the temperature of a gas of atoms is below a critical temperature, a large fraction of these atoms is collapsed into the ground state~\cite{einstein}. At this temperature, bosons undergo a quantum phase transition and they turn into a BEC: an object with coherent wave-like properties in which every atom is in the ground state. This behavior is a direct consequence of quantum statistics, that is, the thermal de Broglie wavelength is of the order of the separation of the particles, and thus their indistinguishability becomes crucial~\cite{dalfovo}. 

Although BEC's had been observed in superconducting and superfluid systems, 
experimental tests with dilute atomic gases was not achieved until 1995 by 
Anderson \emph{et al}~\cite{AEM95} using vapours of rubidium atoms in a 
magneto-optic trap. Almost at the same time, Davis \emph{et al}~\cite{DMA95} 
observed the same phenomenon but using sodium atoms instead of rubidium atoms and Bradley \emph{et al}~\cite{bradley} shown evidences for BEC of a gas of spin-polarized $^{7}Li$. The BEC in a gas of lithium atoms with effective attractive interactions was confirmed~\cite{bradley2}. 
This was a milestone in experimental physics and since then several systems 
have been used to study BEC's and their coherence properties. In 1996, Myatt 
\emph{et al}~\cite{MBG97} created two different condensates in the same trap, 
which corresponds to two different spins states of $^{87}$Rb. These two spin states or species consist of two hyperfine sublevels of $^{87}$Rb, $\vert F,M_F \rangle = \vert 1,-1 \rangle$ and $\vert 2,2 \rangle$,  while in other experiments~\cite{hall1,hall2,GS99} they consider the sublevels $\vert 1,-1 \rangle$ and $\vert 2,1 \rangle$.

Currently, the community recognize that the superconductivity, superfluidity, BEC phenomena and laser light are macroscopic manifestations of quantum behaviour and all of them arise from the macroscopic occupation of a single quantum state. The key ingredients to observe the BEC in diluted alkali gases are the developments of the laser cooling and magneto-optical trapping, and the achievements of the spin-polarized-hydrogen community. The laser cooling was developed by Chu, Cohen-Tannoudji and Phillips~\cite{chu,cohen,phillips}; the alkali atoms are suitable because their optical transitions can be excited by available laser technology and besides they have an energy level structure appropriate for cooling to very low temperatures. It is important to mention that in the conditions of temperature and density to reach BEC, the system would be in the solid phase, then to observe BEC the system must be preserved in a metastable gas phase for a sufficiently long time and this is possible for alkali gases as $^{87}$Rb, $^{23}$Na, and $^{7}$Li. Therefore a typical BEC is a system formed by $10^3$ to $10^6$ atoms, trapped by an harmonic potential with an oscillator length of the order of $10^{-6}$ m, with an average distance between the atoms larger than the range of the interatomic forces and practically with a single parameter, the s-wave scattering length, one can obtain an accurate description.  

We study a two-mode Bose Einstein condensate described in terms of the generators of a $SU(2)$ algebra~\cite{milburn,legget1,cirac1, cirac2}, where the two-modes represent single particle states which can be associated to external or internal degrees of freedom.  The physical system includes independent one and two-body interactions, associated to the atom-atom collisions, the difference in the chemical potentials of the wells and the tunneling amplitude. The parameters associated to each one of the interactions constitutes the parameter space, and we study the stability properties or phase transitions of the ground state of the system when these parameters are varied. In this contribution the quantum phase transitions of the model are established by means of the spin collective states and the catastrophe formalism~\cite{cas1,cas2}. This means to get the locus of points in the parameter space of the model where the ground state of the system has qualitative changes when the parameters of the model are varied, and to this locus of points, we called separatrix. For this reason, the expectation values in the ground state of many observables suffer sudden changes when the parameters of the model take values in different sides of the separatrix. Another purpose is to study the behaviour of the entanglement entropy, the fidelity and the localization in the parameter space of the mentioned two-mode BEC. The entanglement entropy is determined for two partitions of the two-mode BEC, in one of them we separate the system in two boxes~\cite{cirac1, cirac2} and in the other we made the separation in the two hyperfine modes of the atoms~\cite{milburn2}. We show that the entanglement entropy calculation in the boxes partition is strongly correlated to the properties of localization in the phase space of the model, which is given by the evaluation of the second moment of the Husimi function as suggested by Sugita~\cite{sugita1,sugita2}. To compare the probability distributions associated to the variational state and the exact quantum solution we use the fidelity, which in the case of pure states is equivalent to the overlap.  The entanglement entropy for both partitions, the overlap and localization properties of the system get singular values along the separatrix of the two-mode BEC, which indicates where are the phase transitions that remain in the thermodynamical limit, in the parameter space.   

\section{Exact solutions: Dicke and Spin Coherent States}

By means of the Jordan-Schwinger~\cite{schwinger} realization of the components $J_x$, $J_y$, and $J_z$ of the angular momentum operator, a model Hamiltonian that describes a two-mode Bose-Einstein condensate~\cite{milburn,legget1,cirac1} can be written in the following form
\begin{equation}
\hat{H} = \frac{a}{J} \, \hat{J}_z  + \frac{b}{J^2} \, \hat{J}^2_z + \frac{c}{J} \, \hat{J}_x \, ,
\label{original}
\end{equation}
where $J$ denotes the quantum number of the angular momentum operator, besides of indicating the total number of atoms ($N$) in the condensate through the relation $J = N/2$. Because we are interested in taking the thermodynamical limit, the two body interaction is divided by $J$. Finally, to get an intensive quantity the Hamiltonian operator is divided additionally by another factor $J$. The physical meaning of the parameters of the Hamiltonian is the following: the parameter $c$ is related to the single atom tunneling amplitude, $a$ corresponds to the difference in the chemical potentials between the wells, and $b$ represents the atom-atom interaction. 

The stationary states of this model can be obtained analytically for the cases when the parameters $c=0$ and/or $b=0$. In the first case the energies per pairs of particles of the system are given by the expression
\begin{equation}
E_{J, M}\bigl(a,b,0\bigr) = \frac{a}{J} M  + \frac{b}{J^2} M^2 \, ,
\end{equation}
and its corresponding eigenstates are defined by the Dicke states $\vert J, M \rangle$. From here on, we will call energy and energy spectrum to the energy and energy spectrum per pairs of particles. 
This case presents very interesting accidental degeneracy for a finite number of values of the ratio between the parameters $a/b$~\cite{tonel1,cas1}. There is a special value in which all the energy levels are double degenerated except the lowest or highest energy states, and for this case, one can identify the system with a supersymmetric Hamiltonian~\cite{unanyan}. The dynamic behaviour of the energy gap between the ground state and the first excited state for this system has been studied by Tonel et al~\cite{tonel1,cas1}, they do that also with the coupling parameter ($c$) different from zero.

For $b=0$, the Hamiltonian can be diagonalized in terms of an eigenstate of $\hat{J}^2$ and a projection of the angular momentum operator along the direction $\hat n=\Bigl( c/\sqrt{a^2 + c^2},0,a/\sqrt{ a^2 + c^2}\Bigr)$. Thus, the energy eigenvalues are determined by
\begin{equation}
E_{J, M}\bigl(a,0,c\bigr) = \pm \frac{\sqrt{a^2 + c^2}}{J} M  \, ,
\end{equation}
where the corresponding eigenfunctions are generalized Dicke states defined by the expression
\begin{equation}
\vert J, M \rangle_n = \sum_{M\prime} d^j_{M^\prime, M} (\theta_0) \vert J, M^\prime \rangle \, ,
\end{equation}
with $\theta_0= {\hbox{arctan}}\bigl(\pm c/a \bigr)$ and $d^j_{M^\prime, M} (\theta_0)$ the matrix elements of the reduced Wigner rotation matrix~\cite{rose}. It is immediate to recognize that the generalized Dicke state $\vert J, M \rangle_n$ is equivalent to the spin coherent state with parameters $(\theta,\phi) = (-\theta_0, 0)$, when $M=-J$ or $M=J$~\cite{gilmore72,gilmore90}. 

\subsection*{Entanglement properties}

The entanglement is a property of bipartite systems, the concept was introduced by Schr\"{o}dinger in 1935 and it is also mentioned by Einstein-Podolski-Rosen in their manuscript about the incompleteness of quantum mechanics~\cite{epr,schro}. A pure state is entangled when its vector state can not be written as a direct product of pure states of its parts. For this systems the von Neumann entanglement entropy is considered the most basic measure of entanglement to quantify the resources needed to create a given entangled state~\cite{wooters}. Nowadays, the entanglement is recognized as a resource like other physical quantities as energy, or momentum, which can be quantified in terms of the entanglement entropy. Next we will discuss the entanglement properties of the Dicke and spin coherent pure states when they are considered compound systems formed by: the two modes of a two level system or the particles are separated in boxes. For a bipartite pure state system determined by the density matrix $\rho_{A B}$, the entanglement entropy is defined by means of the von Neumann entropy of the reduced density matrices $\rho_A$ or $\rho_B$ as follows
\begin{equation}
S_{E}( \rho_A ) = - \hbox{Tr}{\bigl(\rho_A \ln \rho_A\bigr) }  \, .
\label{neumann}
\end{equation}

A Dicke state can also be written in terms of the eigenstates of two harmonic oscillators using the Schwinger realization of the angular momentum operators~\cite{schwinger}. The harmonic oscillators are characterized by the number of quanta in directions $a$ and $b$, i.e., $\vert J, M \rangle = \vert N-n, n \rangle$, where $N = 2 J$ defines the total number of atoms, $N-n$ denotes the number of particles in the lower state $a$, and $n$ gives the particles in the upper level $b$. 

It is straightforward to construct the total density matrix of a Dicke state, that is $\rho = \vert J, M \rangle \langle J, M  \vert$, by taking the partial trace with respect to the particles in the lower level one finds $\rho_b = \vert n \rangle \langle  n \vert$
that is a reduced density matrix of a pure state. Therefore, if the compound system is described by a Dicke state, the entanglement entropy in the modes partition takes the value $S_E=0$.

%%Figura 1
\begin{figure}[h]
\begin{center}
\scalebox{0.7}{\includegraphics{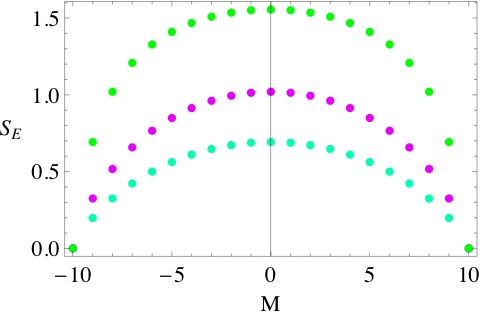}}
\hspace{1cm}
\scalebox{0.7}{\includegraphics{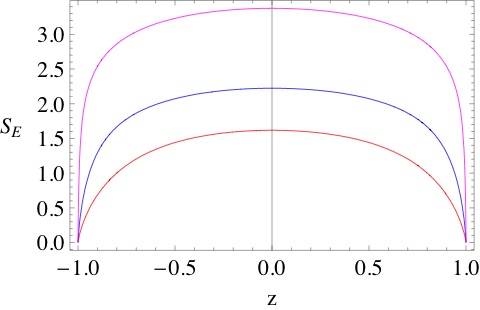}}
\caption{\label{ent-dicke} At the left, the entanglement entropy for Dicke states with $\vert J=10, M \rangle$ with $M=-10,-9,\cdots,9,10$, are shown. It displays the entanglement of one particle (the lowest curve of points), two particles (the middle one curve) and ten particles (the highest curve) with the rest, respectively. The compound system has $N=20$ particles. At the right, The entanglement entropy for spin coherent states with $J=3$, lowest curve, $J=10$, middle plot, and $J=100$, highest curve, are shown. It gives a measure of the entanglement of the particles in the lowest energy state with those occupying the excited energy level.}
\end{center}
\end{figure}

If the system of N atoms is separated in two boxes, one of them with $N_1$ atoms and the other with $N_2$, and satisfying that $N=N_1 + N_2$, the Dicke state can be written as  
\begin{equation}
\vert J, M \rangle = \sum_{\mu} \langle J_1, \mu; J_2, M-\mu \vert J_1 + J_2, M\rangle \vert J_1, \mu \rangle  \, \vert J_2, M-\mu \rangle    ,
\end{equation} 
where $\langle J_1, \mu; J_2, M-\mu \vert J_1 + J_2, M\rangle$ is a Clebsch-Gordan coefficient of the angular momentum theory~\cite{rose}, with $J_1 = N_1/2$, 
$J_2=N_2/2$, such that $\vec J= \vec{J}_1+ \vec{J}_2$.
The Clebsch-Gordan coefficient is a stretched one. Taking the partial trace of the density matrix of a Dicke state with respect to the number of particles in the first box we get the following expression for the reduced density matrix of $N_2$ particles
\begin{equation}
\rho^{(1)}_{m_2^\prime, m_2}(J,M)  =  \frac{{2 J_1 \choose J_1 + m_2} {2 J_2 \choose J_2 + M- m_2}}{ {2 J \choose J + M}} \delta_{m_2^\prime, m_2}    ,
\end{equation} 
where we have substituted the explicit expression for the stretched Clebsch-Gordan coefficient. Thus one gets a diagonal reduced density matrix through which it is immediate to evaluate the entanglement entropy. The results for a Dicke state with $J=10$ for a decomposition in $(N_1,N_2) = (1,19), (2,18)$ and $(10,10)$ subsets are displayed in the left part of Fig.~\ref{ent-dicke}. It is clearly seen that for the lowest or highest value of the projection of the angular momentum one has a pure state and the $S_E=0$. 

%%Figura 1
%\begin{figure}[h]
%\begin{center}
%\scalebox{1.0}{\includegraphics{fig01-entropia_dicke.eps}}
%\caption{\label{ent-dicke} The entanglement entropy for Dicke states with $\vert J=10, M \rangle$ with $M=-10,-9,\cdots,9,10$, are shown. It displays the entanglement of one particle (the lowest curve of points), two particles (the middle one curve) and ten particles (the highest curve) with the rest, respectively. The compound system has $N=20$ particles.}
%\end{center}
%\end{figure}

The spin coherent state of $N$ particles distributed into two modes: $a$ the lowest energy level and $b$ the excited energy level can be written as~\cite{bohr}
	\begin{eqnarray}
		\vert\zeta,\,N\rangle & = &\frac{1}{\sqrt{N!(1+|\zeta|^{2})^{N}}}
		\left(\hat{a}^{\dagger}+\zeta\hat{b}^{\dagger}\right)^{N}
		\vert 0,\,0\rangle
		\label{cstate} \\
		&= & \frac{1}{(1+|\zeta|^{2})^{N/2}}
		\sum_{k=0}^{N}{N\choose k}^{1/2}\zeta^{k}\vert N-k,\,
		k\rangle \ , \nonumber
	\end{eqnarray}
where the state $\vert N-k,\, k\rangle$ is a two dimensional harmonic oscillator state,  the complex number $\zeta$ is usually parametrized in terms of $(\theta,\phi)$ the coordinates of point in a unit radius sphere, i.e., $\zeta=e^{i\phi}\tan(\theta/2)$.
Then the reduced density operator in the first system, i.e., the set of
particles occupying the level $a$, is diagonal and it has the form 
	\begin{eqnarray}
		\rho^{(a)}_{n_1,n_2} &=& \delta_{n_1,\, n_2} \,{N\choose n_1}
		\frac{1}{(1+|\zeta|^{2})^{N}}
		(\vert\zeta\vert^2)^{n_1} \nonumber \\
		&\equiv& \delta_{n_1,\, n_2} \, {\cal P}(n_1)  \ ,
	\end{eqnarray}
where ${\cal P}(n)$ denotes the probability of finding $n$ particles in the upper level in the coherent state. This probability can be identified with a binomial distribution 
where the probability of success is $p=\frac{|\zeta|^{2}}{1+|\zeta|^{2}} = \frac{1-z}{2}$, and $q=1-p$, with $z= \cos\theta$. It is well known that for a binomial distribution, the average number of particles in the upper level $b$ is given by $\langle n\rangle=N \, p$ with dispersion $(\Delta n)^{2}= N \, p \, q$. For $p=q=1/2$ the probability of finding $n$ excited atoms in the system is equal to the one associated to the {\it bonding} state defined in~\cite{milburn2}, which is the eigenstate of the Hamiltonian for the case $a=b=0$ and then it corresponds to a coherent state localized in the equator $(\theta=\pi/2)$ of the unitary sphere. 

%Figura 2
%\begin{figure}[h]
%\begin{center}
%\scalebox{1.0}{\includegraphics{fig02-entropia_coherente.eps}}
%\caption{\label{ent-coh} The entanglement entropy for spin coherent states with $J=3$, lowest curve, $J=10$, middle plot, and $J=100$, highest curve, are shown. It gives a measure of the entanglement of the particles in the lowest energy state with those occupying the excited energy level.}
%\end{center}
%\end{figure}

By means of (\ref{neumann}) it is straightforward to calculate the entanglement entropy of a coherent state in the mode partition that is the entanglement between the atoms in the lowest energy level with those in the excited one. In the right part of Fig.~(\ref{ent-dicke}) we show the entanglement entropy, in nat units, associated to compound systems constituted by $6$, $20$, and $200$ atoms, for $-1\leq z \leq 1$, with $z$ a variable of the spin coherent state. The maxima is reached when the particles are localized into the equator of the Bloch sphere and for $N=100$ particles the entanglement entropy takes a value close to $3.4$ nats. It is straightforward to prove that for $N \rightarrow \infty$ the maximum value of the entanglement entropy $S_E \rightarrow \frac{1}{2} \bigl(1 + \ln{N} + \ln{(\pi/2)}\bigr)$.

The spin coherent state can also be written in terms of the SU(2) generators as~\cite{gilmore72, gilmore90}   
	\[
		\vert\zeta; J \rangle=\frac{1}{\left(1+|\zeta|^{2}\right)^{J}}
		\exp{\bigl( \zeta \, J_+ \bigr)} \vert J, -J \rangle ,
	\]
where the state $\vert J, -J \rangle$ is constituted by the tensorial product of $N$ particles with spin one half. This state is equivalent to the ansatz used in~\cite{legget2}, except at most for an overall phase, with the identification of the zenithal angle: $\theta/2= \theta_L$ of that manuscript. Using the last expression, the spin coherent state, in the boxes partition, can be written in the following form $\vert\zeta; J \rangle \equiv \vert\zeta; J_1 \rangle \otimes \vert\zeta; J_2 \rangle$, with the eigenvalues of the angular momentum operators related by $J = J_1 + J_2$. Therefore one can immediately conclude that the entanglement entropy of the spin coherent state in the boxes partition is zero because the state is separable.

\subsection*{Second Moment of the Q-function}

The Husimi or Q-function~\cite{husimi} gives simply the probability distribution of finding the spin coherent state $\xi$ into the state defined by the density operator of the considered system
	\begin{equation}
	Q_{\rho}( J, \xi ) = \frac{2 J +1}{4 \pi} \, 
	\left\langle \, \xi;J  \vert \rho \vert \xi;J \, \right\rangle \, ,
	\end{equation}
where $\xi= e^{i \phi} \,\tan{\theta/2}$ and the first factor normalizes the Q-function in the complex space $\xi$ or in the unitary sphere defined by the variables $(\theta,\phi)$. Then if the system is determined by a Dicke state the density operator has the form $\rho=\vert J, M \rangle \langle J, M \vert$ and the  Q-function is given by the occupation probability distribution of the two level system, 
	\[
	Q_{\,\vert J,M\rangle}( J, z )= \frac{2 J +1}{4 \pi} \, \,
	{ 2 J \choose J + M} \Bigl(\frac{1 + z}{2}\Bigr)^{J-M}
		\, \Bigl(\frac{1 - z}{2}\Bigr)^{J+M} \, ,
	\]
where we remind you that $\vert \xi \vert^2 = \frac{1-z}{1+z}$. 

\begin{figure}[h]
\begin{center}
\scalebox{1.0}{\includegraphics{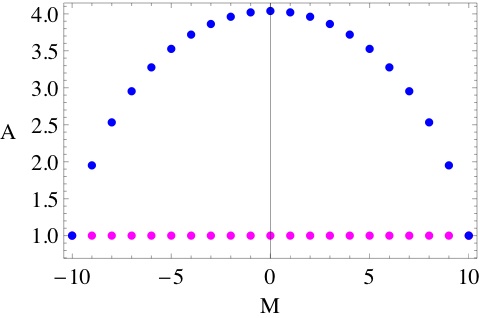}}
\caption{\label{area} The area of phase space, $A$, is displayed for the Husimi distribution of the  Dicke state  $\vert J=10, \,  M \rangle$, for the different projections of the angular momentum, $M=-10,-9, \cdots, 9, 10$. The area of the Husimi function of the spin coherent state is also shown and occupies the minimum area, a constant equal to one. }
\end{center}
\end{figure}

The second moment of the Husimi function has similar properties to the Wehrl entropy, as it is suggested by Sugita~\cite{sugita1,sugita2, wehrl}, and its inverse, as other quasidistribution probabilities, is related with the area of the phase space occupied by the studied system. The second moment is defined by the expression
	\begin{equation}
	M^{(2)}_{Q}( J, M ) = \frac{4 J +1}{4 \pi} \, 
	\int^{\pi}_{0}  \int^{2 \pi}_{0} \sin{\theta}  
	\left\langle \, \theta, \phi  \vert \rho \vert\theta, \phi 
	 \right\rangle^2 \, d \theta \, d\phi\, ,
	\label{segundo}
	\end{equation}
where there is an extra factor depending on the eigenvalue $J$ in the definition of the second moment to guarantee that for a spin coherent state one gets that $M^{(2)}_{ \, \vert \zeta; J \rangle}$ is a constant equal to one~\cite{sugita2}. By means of the expression of the occupation probability in terms of the Wigner's D function and the series of Clebsch-Gordan expression~\cite{rose}, one can obtain the second moment of a Dicke state:  
	\begin{eqnarray}
	M^{(2)}_{Q}( J, M ) &=& \langle J, M; J, M \vert 2 J, 2 M \rangle^2 \,  , \\
	&=& \Biggl(\frac{(2 J)\,!}{(J+M) \, ! \, (J-M) \, !}\Biggr)^2 
	\frac{(2 J + 2 M)\, ! \,
	(2 J - 2 M)\, !}{ (4 J )\, !} \, .
	\end{eqnarray}
Thus the inverse of the second moment for a Dicke state can not be arbitrarily large but at the same time it can not be smaller than $1$. In Fig.~\ref{area}, the areas of phase space occupied for the Husimi functions of a Dicke state and a spin coherent state are shown for $N=20$ particles. For the lowest and highest projection of the angular momentum eigenvalues the Husimi distributions of the Dicke states occupy the same area in phase space that the spin coherent states. 

%Figura 2

\section{General Hamiltonian}

In the previous section we have discussed the analytic solutions (Dicke and spin coherent states) of the model Hamiltonian of a two-mode Bose-Einstein condensate and in particular we have also presented their properties of  entanglement, for the modes and boxes partitions, and localization. Now, we are going to consider the model Hamiltonian~(\ref{original}) when the parameter $b \neq 0$, thus one has a  renormalized Hamiltonian with two essential parameters $(r_a=a/(|b|\,(2J-1), r_c=c/|b|\,(2J-1))$ i.e.,
	\begin{equation}
		\frac{\hat{H}}{|b|\,(2J-1)}=\frac{r_{a}}{J}\, \hat{J}_{z}+
		\frac{\hbox {sign}(b)}{J\,(2 J-1)}\, \hat{J}_{z}^{2}
		+\frac{r_{c}}{J}\, \hat{J}_{x}\, \equiv H_{R,A} \, ,
	\label{gen}
	\end{equation}
where we have defined the function 
	\[
	\hbox {sign}(b) = \left\{
		\begin{array}{ll}
		  \ \ 1  & \hbox{if} \  b > 0 \ , \\
		-1  & \hbox{if} \  b < 0 \ . 
		\end{array} \right.
	\]
The Hamiltonian (\ref{gen}), except for the renormalization of the parameters, has been called the canonical Josephson Hamiltonian which can represent two single particle states separated spatially (external Josephson effect) or characterized by different internal quantum numbers (internal Josephson effect)~\cite{milburn2}. The Hamiltonian can also represent a two-sites version of the Bose-Hubbard model, which describes the dynamics of two species of bosonic atoms moving in an optical lattice~\cite{zoller}. 

According to the interpretation of the different parameters given in Section 2, we are going to study two types of Hamiltonians, one for attractive interactions ($b<0$) between the atoms, $H_A$, and the other for repulsive ones ($b>0$), $H_R$. 

These Hamiltonians can be diagonalized by means of the Dicke basis states either in the two mode realization or in terms of the eigenvalues of the squared of the angular momentum operator $\hat{J}^2$ and its projection $\hat{J}_z$. Thus one proposes that
	\begin{equation}
         \vert \psi^{(k)} \, \rangle = \sum^{N}_{n=0} \, c^{(k)}_n \
	\vert N-n,n \rangle \, ,	
	\end{equation}
where the coefficients $ c^{(k)}_n$ are obtained from the diagonalization of the Hamiltonian, the upper label takes the values $k=0,1,2,\cdots $, denoting the lowest energy state, the first excited state, and so on. 

\subsection{Case $r_a=0$}

This corresponds to the description of a two mode BEC where the chemical potentials of the two wells are equal, meaning that the single particle energies and the scattering lengths for the collisions between the atoms of the different modes of the condensate are identical~\cite{legget2}.

%Figura 3
\begin{figure}[h]
\begin{center}
\scalebox{0.5}{\includegraphics{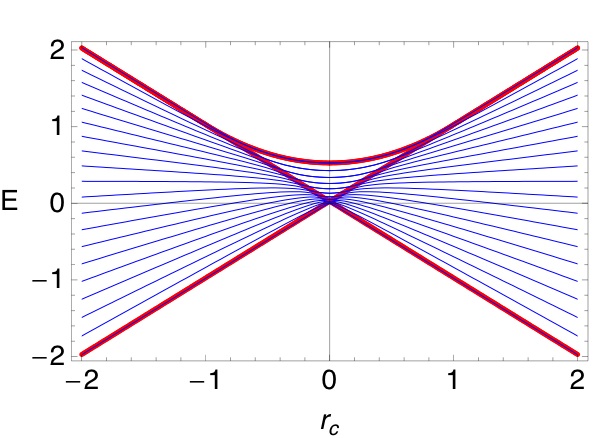}}
\hspace{1cm}
\scalebox{0.5}{\includegraphics{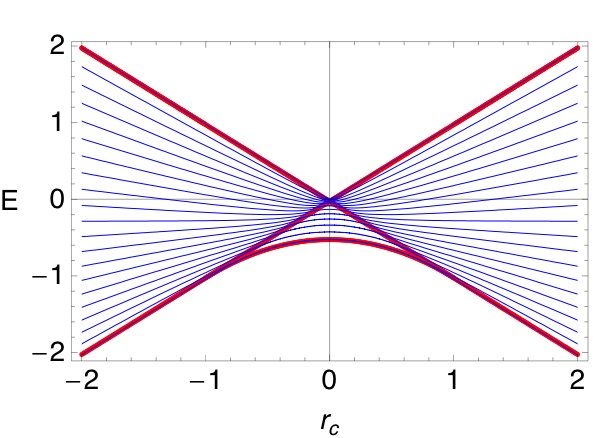}}
\caption{\label{spectrum} The energy spectra of the two-mode BEC, at the left $H_R$ and at the right $H_A$, as functions of the parameter $r_c$ are shown. We are considering the case of $N=20$ particles in the system. The corresponding semiclassical energies are displayed by continuous lines.}
\end{center}
\end{figure}

The energy spectrum of the two-mode BEC of $20$ particles is shown in Fig.~(\ref{spectrum}). The energy spectrum as function of $r_c$ is separated in two regions, one has double degeneration and the other has not. Together with the spectrum of the system we are displaying the corresponding semiclassical energies which are indicated by dark lines. The semiclassical energies are obtained by means of the expectation value of the Hamiltonian $H_{R,A}$ with respect to the variational function constituted by the spin collective coherent states, where the parameters of the test function are determined through a standard minimization procedure~\cite{citlali}. Notice that the energy spectrum of the Hamiltonian $H_R$ is the negative one of the energy spectrum of $H_A$. The degeneracy is due to the symmetry of the Hamiltonian matrix under the interchange of the projection of the angular momentum $M \rightarrow -M$. This suggests us to take symmetric and antisymmetric linear combinations of Dicke states, i.e.,
	\begin{equation}
	\vert M \pm \rangle = \frac{1}{\sqrt{2 \, ( 1 + \delta_{M,0} \, \delta_{\pm,+})}}\left\{
	\vert J, M \rangle \pm \vert J, -M \rangle \right\} \, , 
	\end{equation}
where the label $M = J, J-1, \cdots, \frac{1}{2}, \hbox{or} \, 0$, depending on the value of the angular momentum quantum number $J$ being a half integer or an integer, respectively. In consequence, for an odd number of particles the basis states have dimension $J + 1/2$ while for an even number of particles, the symmetric basis have dimension $J+1$ and for the antisymmetric one dimension $J$. The use of this states is necessary because in the calculations of some expectation values, it allows to avoid numerical instabilities, when the parameter $b$ takes negative values.

%Figura 4
\begin{figure}[h]
\begin{center}
\scalebox{0.5}{\includegraphics{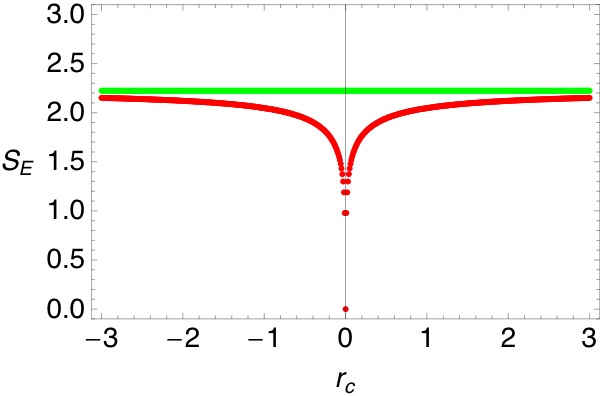}}
\hspace{1cm}
\scalebox{0.5}{\includegraphics{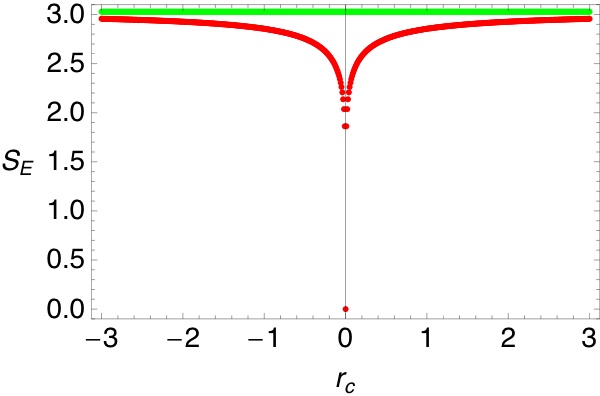}}
\caption{\label{enreda} For $r_a=0$, the entanglement entropies, in the modes partition, of the ground state of the Hamiltonian $H_R$ as functions of the parameter $r_c$ are shown. At the left we consider $N=20$ particles in the system while in the right $N=100$. The corresponding semiclassical entanglement entropies are also displayed and they constitute upper bounds of the entanglement.}
\end{center}
\end{figure}

\noindent
\subsection*{Modes partition}

We evaluate the entanglement entropies for the ground state associated to the repulsive~\cite{milburn2} and attractive Hamiltonians, respectively. 

In Fig.~(\ref{enreda}) we study, for repulsive interactions, the entanglement for two BEC¥s, one with $N=20$ particles and the other with $N=100$ particles. In both cases, the comparison with the entanglement entropy of a spin coherent state is good, outside of the vicinity of $r_c=0$. In this vicinity, it takes the value zero because it corresponds to the entanglement entropy of a Dicke state. Outside that region, the entanglement entropy increases with the number of particles according to the expression $\ln N$, for example if one takes $N=20$ we have $S_E=2.2$ nats while for $N=100$ the result is $S_E=3.0$ nats. Besides the entanglement entropy of the spin coherent state constitutes an upper bound of the quantum result.

\begin{figure}[h]
\begin{center}
\scalebox{0.5}{\includegraphics{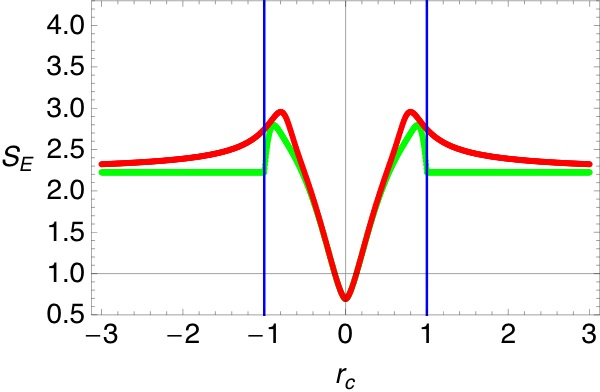}}
\hspace{1cm}
\scalebox{0.5}{\includegraphics{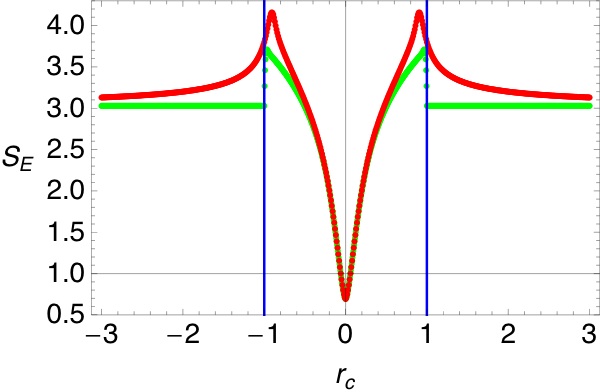}}
\caption{\label{enredaneg} The entanglement entropies in the modes partition of the ground state of the Hamiltonian $H_A$ as functions of the parameter $r_c$ are shown. At the left we consider $N=20$ particles in the system while in the right $N=100$. The corresponding semiclassical entanglement entropies are also displayed in lighter color. The separatrix of the system is shown by the vertical lines at the points $r_c=\pm 1$.}
\end{center}
\end{figure}

The corresponding entanglement entropies, for the ground state of the Hamiltonian $H_A$, are displayed in Fig.~(\ref{enredaneg}). We study again the entanglement for systems with $N=20$ and $N=100$ particles. In both cases, the comparison with the entanglement entropy of a spin coherent state is very good. For this case, the entanglement entropy of the spin coherent state constitutes a lower bound of the quantum result; it increases newly according to the expression $\ln N$. In this figure, the separatrix of the Hamiltonian $H_A$ is shown to indicate the place where the ground state of the system suffers a quantum phase transition.  

In both cases, the behaviour of the entanglement changes abruptly in the separatrix, that is, at the points $r_c= 0$ and $r_c= \pm 1$.

%Figura 5

%Figura 6
\begin{figure}[h]
\begin{center}
\scalebox{0.5}{\includegraphics{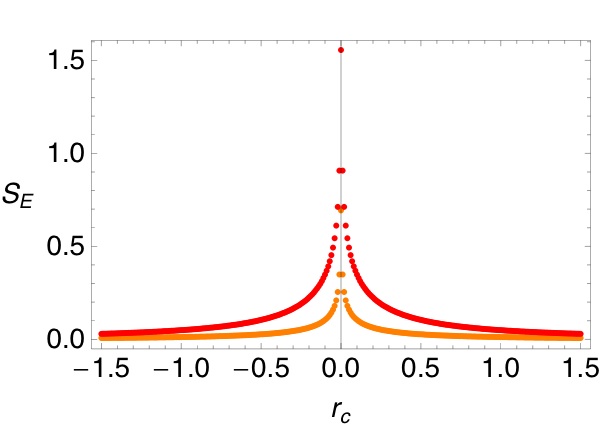}}
\hspace{1cm}
\scalebox{0.5}{\includegraphics{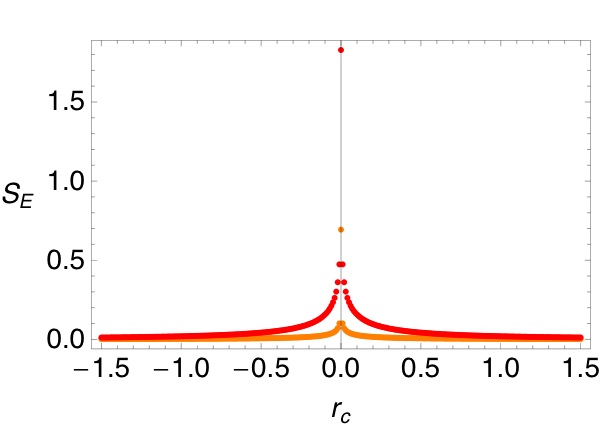}}
\caption{\label{enredafer} The entanglement entropies for two different partitions of the ground state of the Hamiltonian $H_R$ as functions of the parameter $r_c$ are shown, that is the entanglement between one particle and half of particles with the rest. At the left we consider $N=20$ particles in the system while in the right $N=100$. The corresponding semiclassical entanglement entropies are equal to zero because the test function corresponds to a spin coherent state. In each plot, the dark line corresponds to $N_1=N/2$ while the light line to $N_1=1$. }
\end{center}
\end{figure}

\subsection*{Boxes partition}

For the ground state of the Hamiltonian $H_R$, the entanglement entropies are displayed in Fig.~(\ref{enredafer}). The entanglement is shown for systems with $N=20$ and $N=100$ particles with the following partitions: $(N_1=1, N_2=N-N_1)$ and $(N_1=N/2, N_2=N/2)$. The comparison with the entanglement entropy of a spin coherent state is again good only outside of the vicinity of $r_c=0$. In this case, it takes the value zero because it corresponds to the entanglement entropy of a spin coherent state. Then, it is clear that the spin coherent state constitutes a lower bound of the quantum result. Close to the neighborhood $r_c=0$, the entanglement entropy follows the behaviour of the entanglement of a Dicke state.

\begin{figure}[h]
\begin{center}
\scalebox{0.5}{\includegraphics{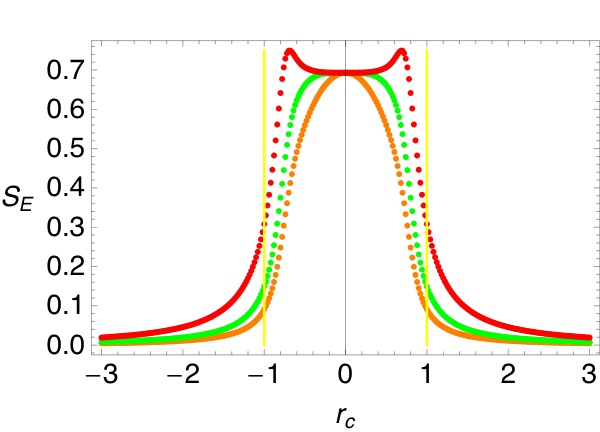}}
\hspace{1cm}
\scalebox{0.5}{\includegraphics{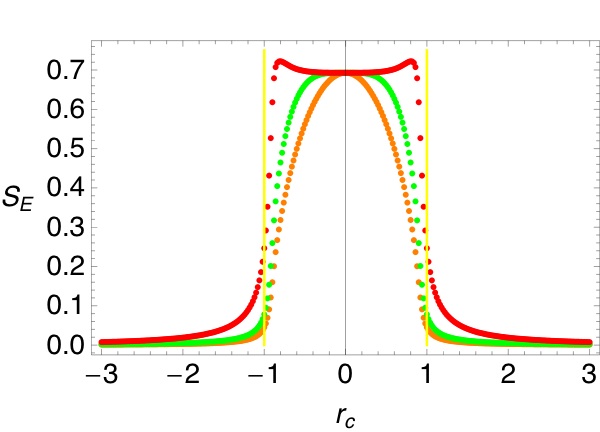}}
\caption{\label{enredanegfer} The entanglement entropies for three different partitions of the ground state of the Hamiltonian $H_A$ as functions of the parameter $r_c$ are shown, that is the entanglement of one, two, and half of the particles with the rest.  At the left we consider $N=20$ particles in the system while in the right $N=100$. The separatrix of the system is shown by the vertical lines at the points $r_c=\pm 1$. The upper curve corresponds to the partition $N_1=N/2$, the next to $N_1=2$ and the lower one to $N_1=1$. }
\end{center}
\end{figure}

For the ground state of the Hamiltonian $H_A$, the entanglement entropies of one, two, and half of the particles with the rest are displayed in Fig.~(\ref{enredanegfer}). We study the entanglement for two systems with $N=20$ and $N=100$ particles; when $-1\leq r_c \leq 1$, the value of the entanglement entropy is the same in both plots. This is due to the fact that the ground state of the system behaves like two qubits with maximum entanglement entropy, that is, they have a value close to $S_E=\ln 2$. Outside this region the ground state of the system has an entanglement entropy of a pure state in this partition. The value of the entanglement entropy changes abruptly in the separatrix, that is at the points $r_c= \pm 1$.  

%Figura 7

\subsection*{Husimi function}

The Q-function is a nonnegative and normalized probability distribution, and is useful to study the quantum correlations. One can conclude that the Q-function for the ground state of condensate with repulsive interactions between the atoms is unimodal while the corresponding Q-function for the condensate with attractive interactions is bimodal. This bimodal behaviour is happening only in the region $-1<r_c<1$; outside that region the Q-function is unimodal~\cite{citlali}. 

\begin{figure}[h]
\begin{center}
\scalebox{0.5}{\includegraphics{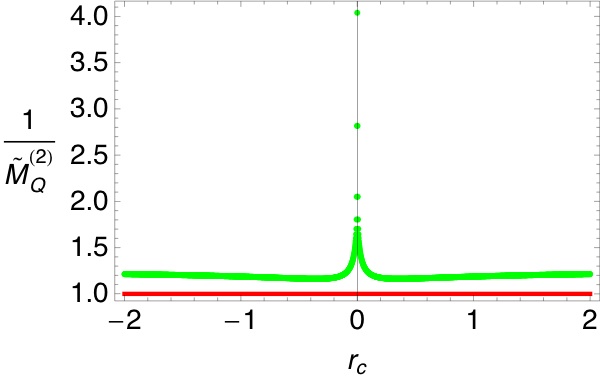}}
\hspace{1cm}
\scalebox{0.5}{\includegraphics{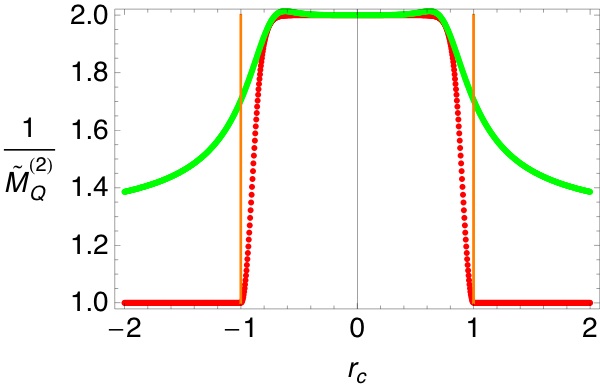}}
\caption{\label{area1} The areas of phase space occupied by the Husimi distributions of the ground state of the two-mode BEC as functions of $r_c$ are shown. At the left for the case of repulsive interactions between the atoms while at the right for attractive interactions. The area of phase space occupied by the coherent state constitutes the lower bound. In the region $-1\leq r_c \leq 1$, the symmetric combination of spin coherent states yields an  area very close to the area of the exact quantum state.}
\end{center}
\end{figure}

The Wehrl entropy attains its minimum for the spin coherent states and has been used to describe complexity of pure states. The second moment of the Husimi distribution has been proposed as a measure of complexity of quantum states because it gives equivalent information than the Wehrl entropy. The inverse represents the effective area or volume occupied by the distribution. It is known that the probability ($p_k$) of finding an eigenvalue of an observable can be used to get the information entropy and their moments distribution, $M_i= \sum_k{p^i_k}$, are measures of delocalization with respect to the basis expansion. In particular the inverse of the second moment is called the number of principal components. However this measure is basis dependent while the inverse of the second moment of the Husimi distribution is not. Thus, the inverse of the second moment of the Q-function constitutes a first measure of delocalization and the spin coherent states have the least delocalized Husimi functions~\cite{karol}. The explicit expressions used for the calculation of the second moment for the Husimi function are presented in the Appendix. This can be seen in both panels of the Fig.~(\ref{area1}) where the delocalization behaviour of the Husimi functions of the ground states of the Hamiltonians $H_R$ and $H_A$ are displayed for a system of $N=20$ particles. From these results, one concludes that the two-mode BEC with attractive interactions between the atoms is more delocalized than the case of repulsive interactions, except when $r_c\approx 0$. In this situation the ground state is described by a Dicke state or very small combination of Dicke states.

%Figura 8

\subsection{Case $r_a \neq 0$ and $r_c \neq 0$}

In this case there are more freedom in the possible values for the chemical potentials of the two wells together with the scattering lengths associated to the collisions between the particles in different modes.

\begin{figure}[h]
\begin{center}
\scalebox{0.5}{\includegraphics{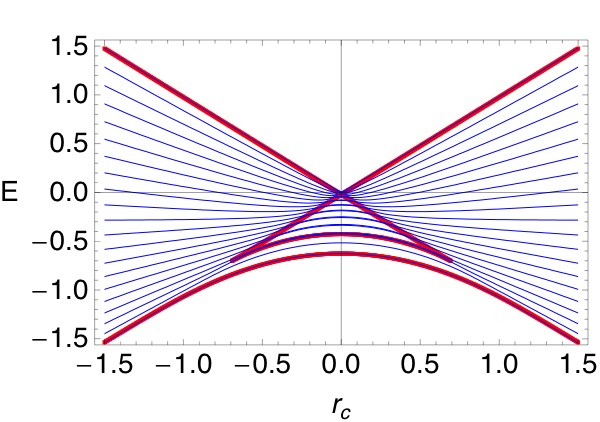}}
\hspace{1cm}
\scalebox{0.5}{\includegraphics{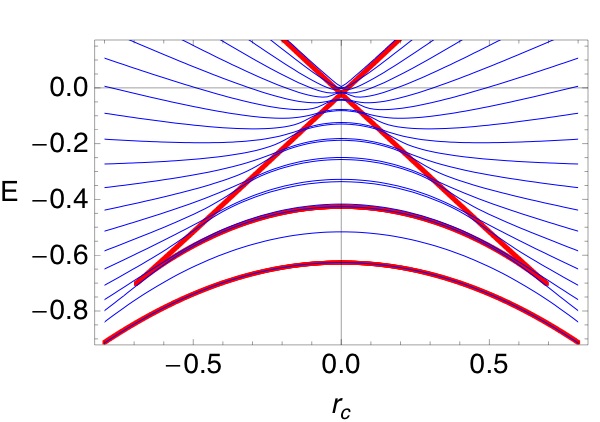}}
\caption{\label{sp3d1} In the left, we display the energy spectrum of the Hamiltonian $H_A$ as function of $r_c$, for a constant parameter $r_a=0.1$. In the right, an amplification of the energy spectrum is shown. The system has $N=20$ particles. The dark continuous lines exhibit the energy surfaces at the critical points.}
\end{center}
\end{figure}

In Fig.~(\ref{sp3d1}) the energy spectrum of the Hamiltonians $H_R$ is shown together with the semiclassical approximation or energy surfaces evaluated at the critical points. One can clearly see that the semiclassical energies separate the spectra into regions where the energy levels present quasi-degeneracy and regions where the energy levels are clearly separated. We consider a system of $N=20$ particles, use a fix value of $r_a=0.1$, and present the spectra as a function of the parameter $r_c$. Besides, the ground state of the Hamiltonian is very well reproduced by the spin coherent state. Here it is immediate to see the effect of the laser interaction by comparing the energy spectra with the one presented in Fig.~(\ref{spectrum}). The interaction breaks the degeneracy of the ground and first excited states. Furthermore the rest of the levels in that region are only quasi-degenerated.

Figura 9

%Figura 10
\begin{figure}[h]
\begin{center}
\scalebox{0.5}{\includegraphics{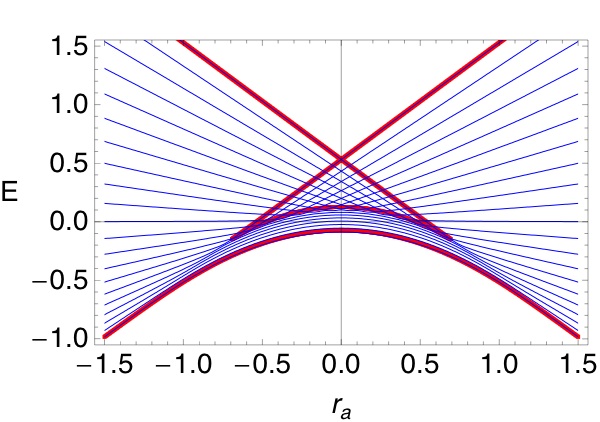}}
\hspace{1cm}
\scalebox{0.5}{\includegraphics{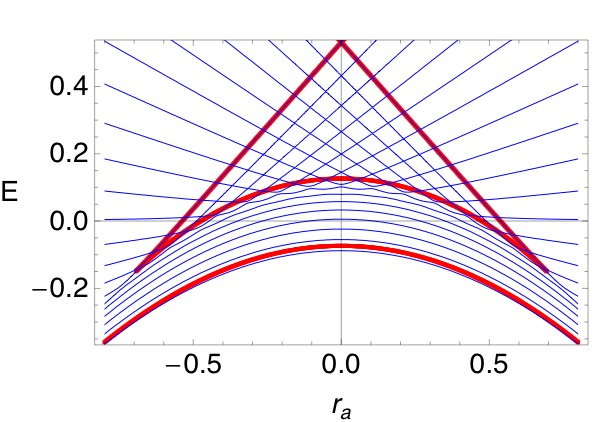}}
\caption{\label{sp3d2} In the left, we display the energy spectrum of the Hamiltonian $H_R$ as function of $r_a$, for a constant parameter $r_c=0.1$. In the right, an amplification of the energy spectrum is shown. The system has $N=20$ particles. The dark continuous lines exhibit the energy surfaces at the critical points.}
\end{center}
\end{figure}

In Fig.~(\ref{sp3d2}) the energy spectrum of the Hamiltonians $H_A$ is shown together with the semiclassical approximation or energy surfaces evaluated at the critical points. The semiclassical energies separate again the spectra into regions where the energy levels present crossings, regions where there are anticrossings, and finally where the energy levels are clearly separated. We consider a system of $N=20$ particles, use $r_c=0.1$, and show the spectra as a function of the parameter $r_a$. The ground state of the Hamiltonian is also very well reproduced by the spin coherent state. Thus one finds that allowing the tunneling between the wells ($r_c\neq0$), the crossings between the energy levels are destroyed~\cite{cas1,cas2}. The energy spectra of the Hamiltonians $H_R$ is the negative one of the energy spectra of  $H_A$, that is the minimum energy of $H_R$ is the maximum energy of $H_A$ and vice versa. Besides the semiclassical energy that constitutes a lower bound of the energy levels with crossings also indicates where there is a zone of anticrossings.

Now, we are going to study the delocalization, and its properties of entanglement, for two types of partitions, of the ground state of the two-mode BEC. This study will be done in the control parameter space for two situations: one of them when the atoms in the condensate have repulsive interactions and the other when it has attractive interactions.    

The ground state of the two-mode BEC condensate, for repulsive interactions, exhibits first order phase transitions for $r_c=0$ within the range of values of the parameter $-1\leq r_a \leq 1$ because the ground state energy surface as function of $r_a$ and $r_c$ presents a peak at that zone. Something similar is happening for the two mode BEC with attractive interactions, for $r_a=0$ there are first order quantum phase transitions at the range $-1\leq r_c \leq 1$ where the manifold presents a peak. The energy surface for the attractive interactions between the atoms is deeper than the energy surface of the repulsive case.

\begin{figure}[h]
\begin{center}
\scalebox{0.5}{\includegraphics{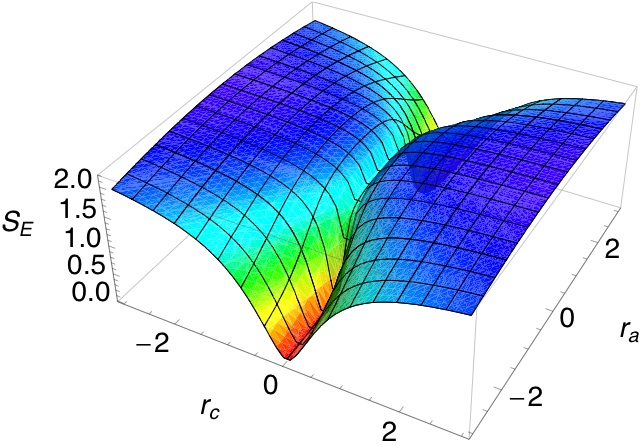}}
\hspace{1cm}
\scalebox{0.5}{\includegraphics{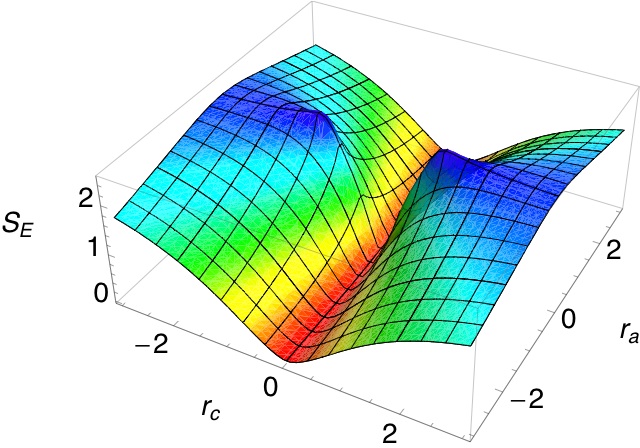}}
\caption{\label{ent3D} The entanglement entropies in the modes partition of the ground state of a two-mode BEC are shown in the parameter space 
$(r_a, r_c)$. At the left for repulsive interactions between the atoms and at the right for attractive ones. We take $N=20$ particles. }
\end{center}
\end{figure}

The entanglement entropy of a compound system depends strongly on the type of decomposition, so in this study we want to know the different behaviours of a two-mode BEC in modes and boxes partitions. 
In Fig.~(\ref{ent3D}), the results in the modes partition for a BEC with repulsive and attractive interactions are shown. 

For the repulsive case the minimum values are obtained when the tunneling amplitude is zero, i.e., $r_c=0$, implying that the lower values of the entanglement occur around that parameter value. One understands this because in that region of parameter values the ground state is dominated by a Dicke state. Relatively far from that region the entanglement entropy takes a value near to $S_E \approx 2$ nats, that corresponds to the value obtained with a spin coherent state for $N=20$ atoms as it can be seen at the right part of Fig.~(\ref{ent-dicke}). 

For the attractive case, the minimum entanglement entropies again are at $r_c=0$, and outside that region the entanglement take larger values close to the points $(r_a=0, r_c=\pm 1)$, even larger that the entanglement for the repulsive case. In general it is immediate to see that the quantum correlations are bigger for the repulsive case than for the attractive one. By making cuts along the line $r_a=0$, one can easily check the results indicated in the Figs.~(\ref{enreda}, \ref{enredaneg}).
 
%Figura 11

\begin{figure}[h]
\begin{center}
\scalebox{0.5}{\includegraphics{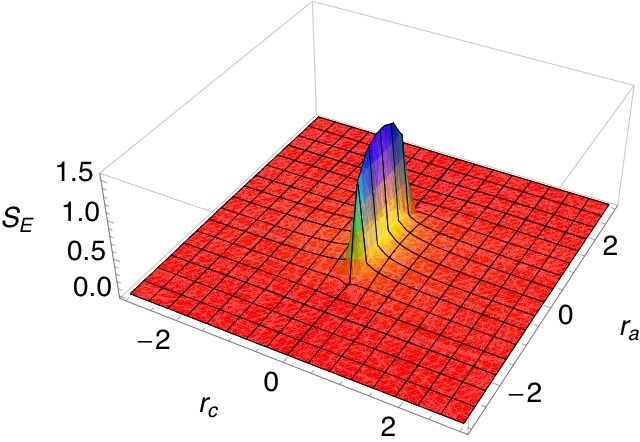}}
\hspace{1cm}
\scalebox{0.5}{\includegraphics{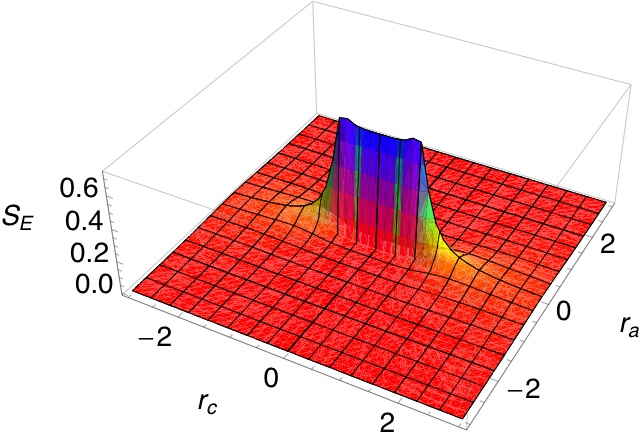}}
\caption{\label{ferent3D} The entanglement entropies in the boxes partition of the ground state of a two-mode BEC are shown in the parameter space 
$(r_a, r_c)$. At the left for repulsive interactions between the atoms and at the right for attractive ones. We study the entanglement entropies of half of the particles with respect to the other half, with $N=20$ particles. }
\end{center}
\end{figure}

In Fig.~(\ref{ferent3D}), the entanglement entropies in the boxes partition for a BEC with repulsive and attractive interactions are shown. For the repulsive case the entanglement entropy is close to zero for almost all values of the control parameter space with the exception of the region near to $r_c \approx 0$ and with chemical potential values of $-1 \leq r_a \leq 1$. The result looks very similar to the entanglement entropy of a Dicke state as one can see in the left part of Fig.~(\ref{ent-dicke}). Thus it can be concluded that the ground state of the system corresponds to Dicke states in that region while outside is very well reproduced by a spin coherent state. 

For the attractive case, the entanglement entropy is near to zero again almost for all the values indicated for the control parameter space with the exception at the region of points close to $r_a \approx 0$ with the tunneling amplitude in the range $-1 \leq r_c \leq 1$.  In that region the entanglement entropy corresponds to the value of the entanglement entropy for a two qubit state with maximum entanglement, as it can be seen in Fig.~(\ref{enredanegfer}). 

Therefore the entanglement entropy for the boxes partition shows that: (i) the entanglement is near to zero in a large region of the control parameter space where the ground state is very well reproduced by a spin coherent state; (ii) for the region where the entanglement is different from zero, the ground state is dominated by a Dicke state or a symmetric combination of two Dicke states.

%Figura 12

%Figura 13
\begin{figure}[h]
\begin{center}
\scalebox{0.5}{\includegraphics{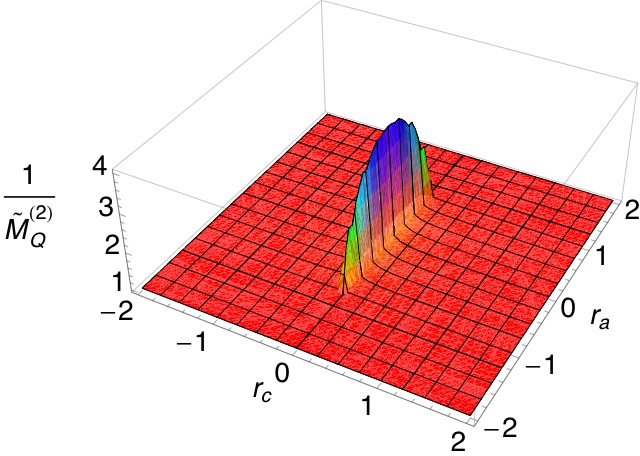}}
\hspace{1cm}
\scalebox{0.5}{\includegraphics{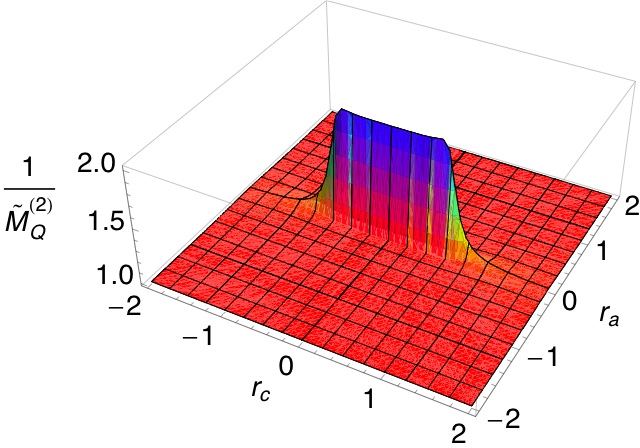}}
\caption{\label{vol3D} The area of phase space occupied by the Husimi distribution of the ground state of a two-mode BEC as function of the control parameter space $(r_a, r_c)$. At the left for repulsive interactions between the atoms and at the right for attractive ones. We use $N=20$ particles. }
\end{center}
\end{figure}

To measure the delocalization we study the second moment of the Husimi distribution for the ground state of the two-mode BEC for repulsive and attractive interactions between the atoms. From the results indicated in Fig.~(\ref{vol3D}), it is immediate that are strongly correlated to the results obtained for the entanglement entropies in the boxes partition for the corresponding ground state. The results show that the repulsive condensate is more delocalized than an attractive one. 

\section{Semiclassical study}    

First of all, one takes the expectation value of the model Hamiltonian (\ref{gen}) with respect to the spin coherent states, getting a function of variables $(\theta, \phi)$ and parameters $(r_a, r_c)$. We call this function energy surface and it is given by
	\begin{equation}
	E(\theta,\phi,r_a,r_c) = -r_a \cos\theta + \frac{\hbox {sign}(b)}{2}\, \cos^2\theta 
	+ r_c \sin\theta \cos\phi +  \frac{\hbox {sign}(b)}{2(2 J -1)} \equiv E_{R,A} \, ,
	\end{equation}
where in the last equality the energy surface is denoted by $E_R$ for the case $b>0$ and $E_A$ when $b<0$. Thus, one gets the critical points of the energy surface by solving the following system of algebraic equations  
	\begin{eqnarray}
	r_a \sin\theta - \hbox{sign}(b) \cos\theta\sin\theta + r_c \cos\theta\cos\phi &=&0 
	\, ,\nonumber \\
	-r_c \sin\theta \sin \phi & =& 0 \, .
	\label{criticos}
	\end{eqnarray} 
For $r_c=0$, the energy surface is $\phi$-unstable and the critical points are given by $\theta_{c_{\, 1}}=0$, $\theta_{c_{\, 2}}=\pi$, and $\theta_{c_{\, 3}}=\arccos(\hbox{sign}(b)\, r_a)$. The energy surfaces evaluated at these points are
	\begin{eqnarray}
	E\vert_{\theta_{c_{\, 1}}} &=& -r_a + \frac{\hbox{sign}(b)}{2}
	\Biggl( 1 + \frac{1}{2 J -1}\Biggr) \, ,\nonumber \\
	E\vert_{\theta_{c_{\, 2}}} &=& \quad r_a + \frac{\hbox{sign}(b)}{2}
	\Biggl( 1 + \frac{1}{2 J -1} \Biggr) \, ,\nonumber \\
	E\vert_{\theta_{c_{\, 3}}} &=& \frac{\hbox{sign}(b)}{2} 
	\Biggl(- r_a + \frac{1}{2 J -1}\Biggr) \, .
	\end{eqnarray} 
For $r_c \neq 0$, the critical points take the values $\phi_{c_1}=0$ and $\phi_{c_2}=\pi$; these values should be substituted in the first row of the expression (\ref{criticos}) to get the corresponding $\theta$ critical points.

Afterwards, the Hessian or stability matrix is calculated at these critical points and by means of the previous values of $\phi_c$ one gets a diagonal matrix whose eigenvalues are
	\begin{eqnarray}
	\lambda_\theta &=& r_a \, z_c - \hbox{sign}(b)\, \Bigl( 2\, z^2_c - 1 \Bigr) 
	- r_c \sqrt{1-z^2_c} \cos\phi_c  \, , \nonumber \\
	\lambda_\phi &=&  - r_c \sqrt{1-z^2_c} \cos\phi_c  \, ,
	\end{eqnarray}
where we do not substitute yet the values of $\phi_c$. Thus the bifurcations sets are obtained when these eigenvalues are zero implying that in the corresponding variable the energy surface can not be approximated around the critical point by a quadratic form.

\begin{figure}[h]
\begin{center}
\scalebox{1.0}{\includegraphics{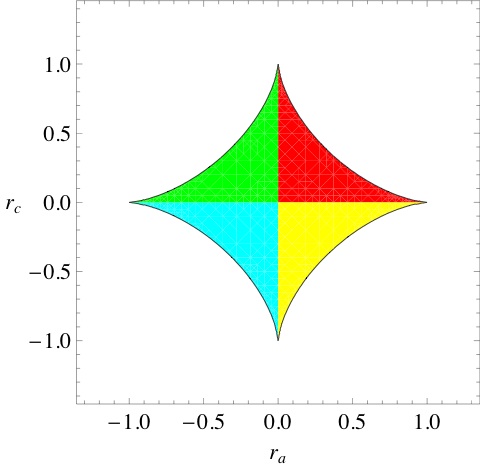}}
\caption{\label{bifurca} The bifurcation set associated to the general Hamiltonian. It has a form of 
a 4 cusps in the parameter space. The critical points for $H_R$ are $\Bigl(r^{1/3}_a, 0\Bigr)$ and $\Bigl(r^{1/3}_a, \pi \Bigr)$ while for $H_A$ one has $\Bigl(-r^{1/3}_a, 0 \Bigr)$ and $\Bigl(-r^{1/3}_a, \pi \Bigr)$.   }
\end{center}
\end{figure}

Next, we follow the procedure indicated in \cite{gilmore} about singularities in mappings to get the bifurcation sets. First of all,  by substituting $\phi_{c_1} = 0$ and $\phi_{c_2}=\pi$ in the first expression of Eqs.~(\ref{criticos}), one get manifolds embedded in a three dimensional space, which are given by
	\begin{equation}
	(z,r_a,r_c)_{0,\pi} = \left(\lambda_1; \lambda_2, \mp \sqrt{1-\lambda^2_1} 
	\Biggl( \frac{\lambda_2}{\lambda_1}- \hbox{sign}(b) \Biggr) \right) \, ,  
	\end{equation}
where the minus sign corresponds to $\phi_{c_1}=0$ and the plus to $\phi_{c_2}=\pi$. Making now the projection to the plane $(r_a, r_c)$ one has an invertible map in general except for
	\begin{equation}
	\lambda_2 = \hbox{sign}(b)\, \lambda^3_1 \, ,
	\label{rc}
	\end{equation}
which let us define a curve in the control parameter space with the localization of all degenerate critical points of $E_R$ and $E_A$, that is 
	\begin{equation}
	\Bigl( r_a \Bigr)^{2/3} + \Bigl( r_c \Bigr)^{2/3} =1 \, .
	\label{sepnew}
	\end{equation}
This bifurcation set yields a contour plot, Fig.~\ref{bifurca}, of 4 cusps separating the control parameter space in open regions where the critical points are not degenerated. The 4 cusps have been produced by two swallowtail singularities, these can be defined as the set of all points $(A,B,C)$ such that the polynomial $z^4 + A z^2 \pm B z + C$ has a multiple root. It can be seen that when $r_c=0$ the range of the parameter $r_a$ is bounded by $-1\leq r_a \leq 1$ and vice versa. Notice that by plotting the critical values of $z$ in the plane $(r_a,r_c)$ one finds that outside the 4 cusps there are two real critical points while in the inside part there are four critical points. On the bifurcation set there are two degenerated critical points.

%Figura 15

We are looking for the general solution of (\ref{criticos}), thus from the second equation one gets that $\phi_{c_1} = 0$ and $\phi_{c_2}=\pi$ and the corresponding critical values of $\theta$ are obtained through the solutions of the quartic equation
	\begin{equation}
	z^4 - 2 \, \hbox {sign}(b)\, r_a z^3 + (r^2_a + r^2_c -1 ) z^2 + 2 \, \hbox {sign}(b)
	\, r_a z - r^2_a = 0  \, , 
	\end{equation}
where $z=\cos \theta$. Notice that it is independent of the sign of the parameter $r_c$. It is straightforward to prove also by means of the expressions (\ref{rc}) and (\ref{sepnew}) that the quartic equation is satisfied  by the bifurcation set. 

For $r_a=0$, the quartic equation is simplified and one finds the following critical points, namely for $z=0$, one has $(\theta_{c},\phi_{c})=(\pi/2, 0)$ and $(\pi/2, \pi)$, while for $z^2 = 1- r^2_c$ the points are given by $(\theta_{c_{\, 3}},\phi_{c_{\, 3}})=(\arcsin({\hbox {sign}(b)\, r_c}),0)$ and $(\theta_{c_{\, 4}},\phi_{c_{\, 4}})=(\arcsin({-\hbox {sign}(b) \, r_c}),\pi)$. The last two points are real only when the condition $-1 \leq r_c \leq 1$ is satisfied. Besides it is straightforward that for $r_c=\pm 1$ the critical points are degenerated in pairs. The corresponding energy surfaces are given by the expressions
	\begin{eqnarray}
	E\vert_{(\pi/2,0 )} &=& \ r_c + \frac{\hbox{sign}(b)}{2(2 J -1)} \, ,\nonumber \\
	E\vert_{(\pi/2,\pi )} &=&  - r_c + \frac{\hbox{sign}(b)}{2(2 J -1)} \, ,\nonumber \\
	E\vert_{(\theta_{c_{\, 3}},\phi_{c_{\, 3}})} &=& \frac{\hbox{sign}(b)}{2} 
	\Biggl(1 + \frac{1}{2 J -1}\Biggr) \equiv E\vert_{(\theta_{c_{\, 4}},\phi_{c_{\, 4}})} \, .
	\end{eqnarray} 

%Figura 16
\begin{figure}[h]
\begin{center}
\scalebox{0.6}{\includegraphics{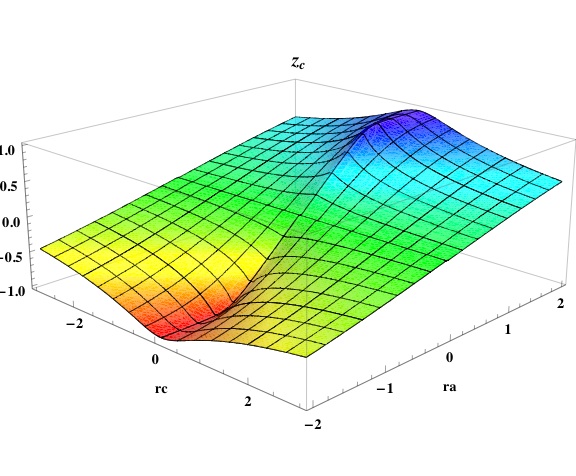}}
\hspace{1cm}
\scalebox{0.6}{\includegraphics{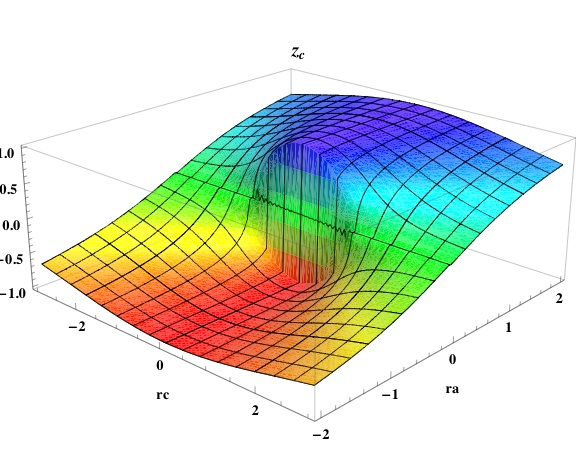}}
\caption{\label{zminima} The $z_c$ minima as function of the control parameter space $(r_a,r_c)$. At the left the $z_c$ minimum corresponds to repulsive interactions between the atoms of the two-mode BEC while at the right to the $z_c$ minimum for attractive interactions.  }
\end{center}
\end{figure}

The solution of the quartic equation gives all the critical points, from them we consider those that yield a minimum energy surface. Thus, the obtained minima $z_c$ are displayed in Fig.~(\ref{zminima}) for the repulsive and attractive interactions between the two-mode BEC. For the repulsive case, the values of the $z_c$ minimum presents a discontinuity along the straight line $r_c=0$ in the region $-1<r_a<1$ while in the attractive one, it is immediate that the $z_c$ changes suddenly along the line $r_a=0$ in the region $-1<r_c<1$, that is when the parameter $r_a$ changes from zero to small positive or negative values.

\section{Exact vs. semiclassical results}

The ground state energies of the two-mode BEC for repulsive and attractive interactions between the atoms have been very well described by the collective spin coherent states as functions of the control parameters $r_a$ and $r_c$. The ground state energies have symmetries under the interchange of $r_c \rightarrow -r_c$ and $r_a \rightarrow -r_a$. 

We compared the semiclassical results of the ground state of the system using cuts: $r_c = 0.5$ and $r_c=1.8$; and  $r_a=0.5$ and $r_a=1.8$. The energy differences were of the order of $10^{-3}$ or lower. If one considers the locus of points associated to the 4 cusps (\ref{sepnew}) and determine the energy spectrum of the two-mode BEC, it is found that the separatrix corresponds to the points $(r_a,r_c)$ where the  differences between the energy levels, i.e., $\Delta E (q) \equiv E(q+1)-E(q)$ with $q=1,2, \cdots, 2 J$, take minima values. In other words, the separatrix identifies the region of anticrossings of the model.  

The exact quantum calculation of the expectation value of the relative population operator ($\hat{J}_z$), divided by $J$, corresponds the negative values of the semiclassical result for $z_c$ minimum. The expectation values are calculated for the repulsive and attractive interactions between the atoms of the two-mode BEC. These expectation values as functions of the control parameter space are indistinguishable to the plots presented in Fig.~(\ref{zminima}). Similar results for the fluctuations of the angular momentum operators are easily calculated and compared with the semiclassical evaluations~\cite{citlali}.  

\begin{figure}[h]
\begin{center}
\scalebox{0.5}{\includegraphics{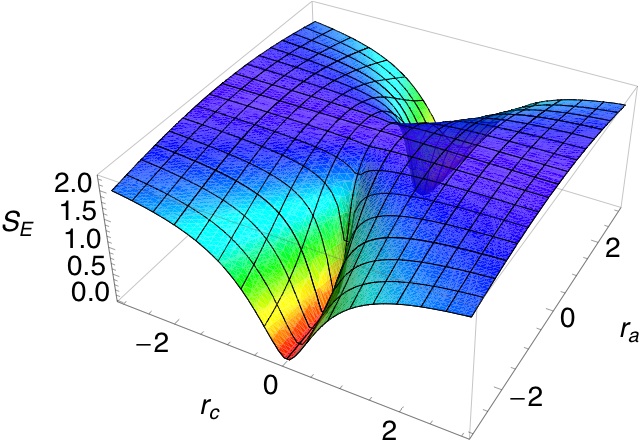}}
\hspace{1cm}
\scalebox{0.55}{\includegraphics{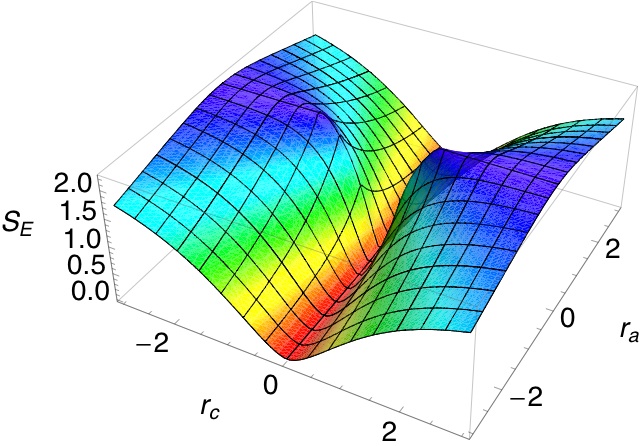}}
\caption{\label{entclas3D} The entanglement entropies in the modes partition of the ground state of a two-mode BEC described with the spin coherent states are shown in the parameter space 
$(r_a, r_c)$. At the left for repulsive interactions between the atoms, and at the right for attractive ones. We take $N=20$ particles. }
\end{center}
\end{figure}

The entanglement entropy in the modes partition for the ground state of the two-mode BEC described by the spin coherent state is given in Fig.~(\ref{entclas3D}). The comparison with the corresponding results for the exact quantum states, for the case of repulsive interactions between the atoms, is remarkable, except in the region of points close to the $r_c=0$. For attractive interactions between the atoms in the two-mode BEC, it seems that the semiclassical results overestimates the entanglement entropy. For the boxes partition, the semiclassical result gives zero value for the entanglement entropy for both cases. 

%Figura 17

\begin{figure}[h]
\begin{center}
\scalebox{0.55}{\includegraphics{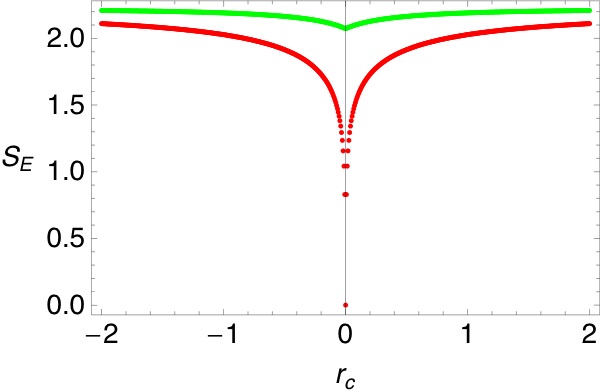}}
\hspace{1cm}
\scalebox{0.55}{\includegraphics{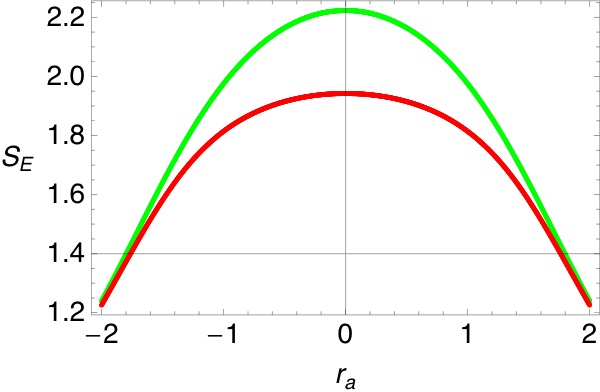}}
\caption{\label{entboscortes1} The entanglement entropies in the modes partition of the ground state of repulsive interactions between the atoms of a two-mode BEC are shown. The variational spin coherent state is an upper bound of the exact calculation. At the left as function of $r_c$ with $r_a=0.5$, and at the right as function of $r_a$ with $r_c=0.5$. We take $N=20$ particles. }
\end{center}
\end{figure}

\begin{figure}[h]
\begin{center}
\scalebox{0.55}{\includegraphics{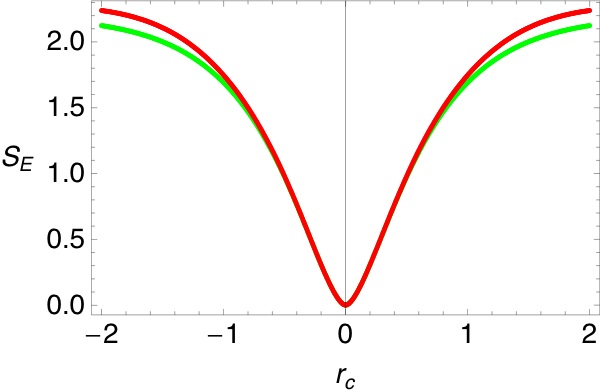}}
\hspace{1cm}
\scalebox{0.55}{\includegraphics{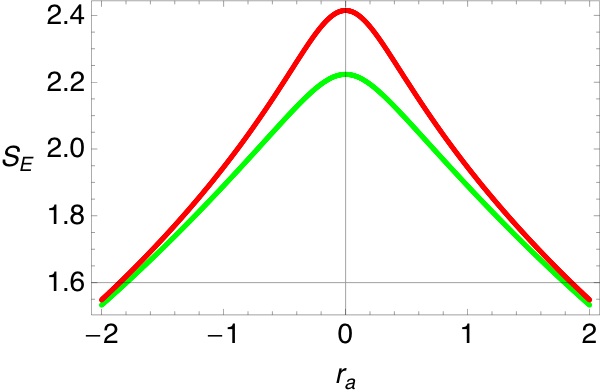}}
\caption{\label{entboscortes2} The entanglement entropies in the modes partition of the ground state of attractive interactions between the atoms of a two-mode BEC are shown. The variational spin coherent state is a lower bound of the exact calculation. At the left as function of $r_c$ with $r_a=0.5$, and at the right as function of $r_a$ with $r_c=1.8$. We take $N=20$ particles. }
\end{center}
\end{figure}

According to the results of the entanglement entropy in the modes partition for the exact and variational results, for repulsive interactions one gets, in the vicinity of $r_c=0$ and in the region $-1<r_a<1$, the largest differences. For this reason, we consider in Fig.~(\ref{entboscortes1}) the entanglement entropy as function of $r_c$ with $r_a=0.5$ and as function of $r_a$ with $r_c=0.5$. The obtained results are in good agreement except for the region close to the line $r_c=0$. Besides for other regions the variational spin coherent state is an upper bound of the exact calculation. For attractive interactions between the atoms of the two-mode condensate the agreement is very good in almost all the regions in the control parameter space. This is confirmed in Fig.~(\ref{entboscortes2}) where the entanglement entropy is plotted as function of $r_c$ with $r_a=0.5$ and as function of $r_a$ for $r_c=1.8$. The comparison is very good and for this case, the variational results is a lower bound of the exact result. These considerations about the entanglement entropy are general in all the regions of the control parameter space. 

%Figura 18

%Figura 19

Finally, we consider the overlap between the exact and variational wave functions of the two-mode BEC for repulsive and attractive interactions between the atoms. For repulsive interactions, the percentage of overlap is close to one hundred except in the region $-1 \leq r_a \leq 1$ in the vicinity of the straight line $r_c=0$. For attractive interactions, the largest disagreements between the exact and the test ground state occur in the neighborhood of the points $r_a=-1$ and $r_a=1$. Anyway in both cases the lowest overlap percentages are of order $50\%$ to $60\%$. We want to enhance that the overlap has been used to study the regions of criticality that define quantum phase transitions~\cite{zanardi}, where they study the fidelity in the Dicke and XY model Hamiltonians.

%Figura 24
\begin{figure}[h]
\begin{center}
\scalebox{0.55}{\includegraphics{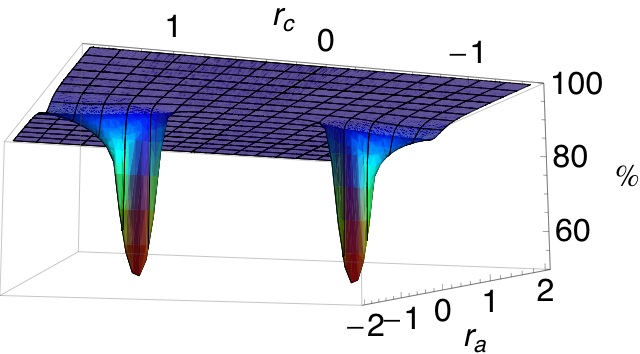}}
\hspace{1cm}
\scalebox{0.55}{\includegraphics{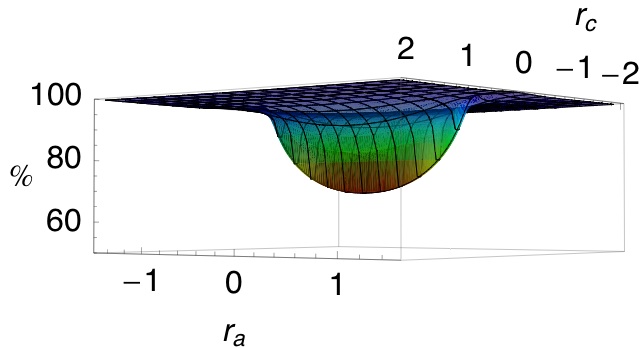}}
\caption{\label{traslape} The overlap between the exact quantum solution for the ground state with the corresponding spin coherent variational state. At the left the overlap is calculated for attractive interaction between the atoms, while at the right for repulsive ones. We take $N=20$ particles. }
\end{center}
\end{figure}

\section{Conclusions}

The ground state properties of a two-mode BEC described by the
intensive Hamiltonian~(\ref{gen}) has been shown. This was done by
means of the exact quantum diagonalization and through a variational spin
coherent state. The studied properties include: the energies,
entanglement entropies in two partitions of the system, the Husimi
function and its second moment. From this, we conclude that the ground state is well
described by a Dicke state, a spin coherent state, or a symmetric
combination of spin coherent states, depending on the region of the
control parameter space. The critical points of the energy surface of
the Hamiltonian let us organize the energy spectrum in regions where
there are degeneracy and anticrossings. The separatrix of the model was determined, which
establishes the stability properties of the ground state of the system and indicates its phase transitions.

For $r_{a}=0$, the ground state of the Hamiltonian
has an atom population equally distributed in the two hyperfine
sublevels, i.e., $\langle {\hat J}_z \rangle=0$, or, if we think in the two-mode as a potential
with two wells, we have the same number of atoms in each well. The
spectrum of repulsive $b>0$, and attractive $b<0$ interactions between the atoms in the BEC are inverted, that is one is the negative of the other. The critical values of the energy surface separate
both energy spectra into regions with and without degeneracy. 

The Husimi distribution of the ground state of the two mode BEC shows a bimodal behaviour for attractive interactions between the atoms approximately within the interval $-1 \leq r_c \leq 1$, while outside the previous region the behaviour is unimodal. The bimodal behaviour occurs in the whole mentioned interval, when the number of particles is very large. For repulsive interactions one gets always a unimodal characteristic for the corresponding Husimi function. 

For attractive interactions between the atoms, we
need to use a symmetric combination of coherent states to get a better
fit. For the modes partition, the entanglement entropy calculated with
the trial state gives an upper bound for the exact result for the
repulsive potential, while for the attractive potential, it constitutes
a lower bound. For the boxes partition, the entanglement entropy for repulsive
interactions is almost zero and independent of the number of atoms outside the vicinity of $r_{c}=0$, while for attractive interactions it is constant and almost equal to the value of a Bell state in the interval $-1 \le r_c \le 1$.

For $r_a \neq 0$ and $r_c \neq 0$, we obtain a similar behaviour, the energy spectrum is divided by the semiclassical energies in regions with quasi-degeneracy and without degeneracy. In general, the ground state of the two-mode BEC is very well described by the spin coherent state in almost all the control parameter space. This good agreement is not happening for repulsive interactions at the locus of points close to the line $r_c=0$,  while for the attractive case this occurs in the vicinity of the points $r_a=-1$ and $r_a=1$. This is proved by the evaluation of the overlap of the spin coherent state with the exact quantum solution. The region of anticrossings of the model is determined by the 4 cusps in the control parameter space, which defines the separatrix.

The entanglement entropies for the boxes and modes partitions of the ground state of the two mode BEC,
were calculated, and the match between our
trial state description with the exact solution is very good in the $(r_{a},\,r_{c})$
control parameter space. The exception is happening, for attractive interactions, in the neighborhood of line $r_{c}=0$,
where the ground state of the two mode BEC corresponds a Dicke state with projection of the angular momentum operator $\hat{J}_z$ equal to zero.

The calculations of the entanglement entropies, the second moment of the Husimi functions, and the distribution of particles in the hyperfine levels take singular values close to the separatrix. The population of atoms for the attractive interactions is mainly distributed in the lower level for $r_a<0$ and they are in the upper level for $r_a>0$. For the repulsive case they are mainly equally distributed into the levels, except in the neighborhood of $r_a=-1$ and $r_a=1$.  

It has been shown that there is a correlation of the entanglement entropy in the boxes partition with the second moment of the Husimi function as it was suggested by Sugita, that is, the entanglement entropy is maximum where the area of the phase space of the Husimi function is maxima, and vice versa. This is very important to have global measures of the entanglement of pure states consisting of two parts. 

In this work, we have been considering the static properties of the two mode BEC, similar studies related with the dynamics of a two mode BEC have been published recently. In 2001, the dynamics of a two mode BEC was studied in the vicinity of the mean field dynamical instabilities~\cite{vardi}, more recently in 2005 the evolution of an arbitrary state have been investigated for the two mode BEC across a wide range of coupling regimes~\cite{TLF05} or in similar type of models~\cite{VPA04}.  Following the procedure used in these contributions together the time dependent variational principle~\cite{KS81}, we will study the dynamic behaviour of the entanglement entropies and second moment of the Husimi function in the control parameter space $(r_{a},\,r_{c})$. By means of the appropriate selection of the variational states, the procedure established in this work can also be applied to other Hamiltonian systems involving two mode approximations like those describing the coupling between atom-molecule BEC modes~\cite{santos}.

\section*{Appendix. Evaluation of the second moment of the Husimi function}

We consider the symmetric superposition of two spin coherent states
	\[
	\vert \zeta_1, \, \zeta_2 , \,J \rangle = {\cal N}( \zeta_1, \zeta_2) 
	\Bigl( \vert \zeta_1, J \rangle  
	+ \vert \zeta_2, J \rangle \Bigr) \, , 
	\]
where 
	\[
	\vert {\cal N}( \zeta_1, \zeta_2)\vert= \sqrt{\frac{(1 + \vert \zeta_1\vert^2)^J \, 
	(1 + \vert \zeta_2\vert^2)^J}{2 \, \Bigl\{ (1 + \vert \zeta_1\vert^2)^J 
	(1 + \vert \zeta_2\vert^2)^J 
	+ \hbox{Re}(1 + \zeta_1^* \zeta_2 )^{2 J} \Bigr\}}} \, ,
	\]
denotes the normalization constant of the state. The second moment of the Husimi function is given by the expression (\ref{segundo}), with the density operator $\rho= \vert \zeta_1, \, \zeta_2 , \,J \rangle \, \langle \zeta_1, \, \zeta_2 , \,J \vert$. For simplicity from here on, we will forget the label of the angular momentum quantum number $J$ in the spin coherent state. Then, to calculate the second moment, we consider $\langle \, \xi  \vert \zeta_1 , \, \zeta_2 \rangle^2$, which is given by the expression
	\begin{equation}
	\langle \, \xi \vert \zeta_1, \, \zeta_2 \rangle^2 = 
	{\cal N}( \zeta_1, \zeta_2)^2 \Bigl( \langle \, \xi \vert \zeta_1 
	\rangle^2 +  2 \langle \, \xi \vert \zeta_1 \rangle \langle \, \xi \vert \zeta_2 \rangle
	+ \langle \, \xi \vert \zeta_2 \rangle^2 \Bigr) \, .
	\label{B1}
	\end{equation}
Remembering the expression of a spin coherent state in terms of the D Wigner matrices, that is
	\[
	\vert \zeta  \rangle = \sum_{M^\prime} D^J_{M^\prime, -J} 
	(\Omega_\zeta) 	\vert J, M^\prime \rangle \, , 
	\]
one can evaluate all the terms appearing in (\ref{B1}), 
	\[
	\langle \, \xi \vert \zeta_k \rangle = \sum_{M} 
	D^{J \, *}_{M, -J} (\Omega) \,  D^J_{M, -J}(\Omega_{\zeta_k})  \, . 
	\]
Substituting the corresponding previous results in (\ref{B1}), the expression for the overlap $\langle \, \xi \vert \zeta_1 ,\, \zeta_2 \rangle^2$, can be written as follows 
	\begin{equation}
	\langle \, \xi \vert \zeta_1, \, \zeta_2 \rangle^2 = {\cal N}( \zeta_1, \zeta_2)^2 
	\, \sum_{\mu} \, D^{2 J \, *}_{\mu, -2 J} (\Omega)
	\, F^{J}_\mu(\Omega_{\zeta_1}, \Omega_{\zeta_2})  \, , 
	\label{B2}
	\end{equation}
where we define the function
	\begin{eqnarray*}
	F^{J}_\mu(\Omega_{\zeta_1}, \Omega_{\zeta_2}) &=&  
	\,  D^{2 J }_{\mu, -2 J}(\Omega_{\zeta_1}) +
	D^{2 J }_{\mu, -2 J}(\Omega_{\zeta_2}) \nonumber \\
	&+&  2 \sum_{M} \langle J, \mu-M, J, M \vert 2 J, \mu \rangle, 
	D^J_{\mu-M, -J}(\Omega_{\zeta_1}) \, D^J_{M, -J}(\Omega_{\zeta_2})  \, , 
	\end{eqnarray*}
where we use the Clebsch-Gordan series together with the orthonormality properties of the Clebsch-Gordan coefficients~\cite{rose}. Thus, taking into account that the solid angles for a spin coherent state only depend on two angles, those that define a point in the unit sphere, the integration can be easily realized when the expression (\ref{B2}) is multiplied by its complex conjugated. Therefore the second moment of the Husimi function is given by
	\[
	M^{(2)}_{Q(\zeta_1,\, \zeta_2}( J, M )= {\cal N}( \zeta_1, \zeta_2)^4 
	\, \vert F^{J}_\mu(\Omega_{\zeta_1}, \Omega_{\zeta_2})\vert^2 \, .
	\]
For the symmetric combination of coherent states, one has to replace $\Omega_{\zeta_1} \rightarrow (-\phi_c,-\theta_c,\phi_c)$ and $\Omega_{\zeta_2} \rightarrow (-\phi_c,-\pi+\theta_c,\phi_c)$.

\section*{Acknowledgments}

We want to thank to J. Hirsch and E. L\'opez Moreno for their remarks and suggestions. This work was supported by CONACyT-Mexico.

\end{document}